\documentclass[12pt,reqno]{amsart}%
\usepackage{graphicx}
\usepackage{amscd}
\usepackage{amsmath}
\usepackage{epsfig}
\usepackage{amsfonts}
\usepackage{amssymb}%
\providecommand{\U}[1]{\protect\rule{.1in}{.1in}}
\numberwithin{equation}{section}
\providecommand{\U}[1]{\protect\rule{.1in}{.1in}}
\providecommand{\U}[1]{\protect\rule{.1in}{.1in}}
\allowdisplaybreaks

\theoremstyle{plain}

\begin{document}
\title[Photon Statistics]{Time-Dependent Photon Statistics in Variable Media}
\author{Sergey~I.~Kryuchkov}
\address{School of Mathematical and Statistical Sciences, Arizona State University,
Tempe, AZ 85287--1804, U.S.A.}
\email{sergeykryuchkov@yahoo.com}
\author{Erwin~Suazo}
\address{School of Mathematical and Statistical Sciences, University of Texas of Rio
Grande Valley, 1201 W. University Drive, Edinburg, TX 78541, U.S.A.}
\email{erwin.suazo@utrgv.edu}
\author{Sergei~K.~Suslov}
\address{School of Mathematical and Statistical Sciences, Arizona State University,
Tempe, AZ 85287--1804, U.S.A.}
\email{sks@asu.edu}
\urladdr{http://hahn.la.asu.edu/\symbol{126}suslov/index.html}
\date{November 24, 2016.}

\begin{abstract}
We find explicit solutions of the Heisenberg equations of motion for a
quadratic Hamiltonian, describing a generic model of variable media, in the
case of multi-parameter squeezed input photon configuration. The corresponding probability
amplitudes and photon statistics are also derived in the Schr\"{o}dinger
picture in an abstract operator setting of the quantum electrodynamics (QED).
Their time evolution is given in terms of solutions of certain Ermakov-type
system. The unitary transformation and an extension of the squeeze/evolution
operator are introduced formally.

\end{abstract}
\maketitle

\section{An Introduction}

Although for most applications in optics, the electromagnetic field can be
treated classically \cite{GlauberNobel}, \cite{GlauberCollect},
\cite{Scully:Zubairy97}, a quantum description is required when quantum limits
are approached and one is interested in the photon statistics of the field
(see \cite{Drummomd90}, \cite{Glauber91}, \cite{Hillery09}, \cite{Hillery84}
and the references therein). A fundamental effect of squeezing, or
simultaneous (counter-)oscillations of variances of the electric and magnetic
field operators, is one of the characteristic features of quantized light
propagating in a variable medium \cite{Bialynicka-Birula87}. Observations of
squeezed states of light, generated by advanced methods of nonlinear optics,
are discussed in \cite{BachorRalph04}, \cite{Breit:Schill:Mlyn97},
\cite{LeonardPaul95}, \cite{LvRay09} (see also the references therein).
Various aspects of the corresponding photon statistics and photon-counting
were studied in detail \cite{Agarwal87}, \cite{AgarwalAdam88}, \cite{Caves81},
\cite{Caves82}, \cite{Dod:Man:Man94}, \cite{Glauber63}, \cite{KimOlivKnight89}%
, \cite{KolesManko08}, \cite{MankoWuensche97}, \cite{MarhicKumar90},
\cite{Marian91}, \cite{Marian92}, \cite{MarianMarianI93},
\cite{MarianMarianII93}, \cite{Stoler70}, \cite{Stoler71}, \cite{Stoler74},
\cite{VenkSatya85}, \cite{VourdasWeiner87}, \cite{Yuen76}.

In spite of the considerable literature on squeezing and photon statistics,
the most general case of time evolution of multi-parameter squeezed input
photons propagating in a variable inhomogeneous medium, to the best of our
knowledge, has never been thoroughly discussed with an exception of
\cite{Agarwal87}, \cite{Dod:Man:Man94}, \cite{KolesManko08}, \cite{Stoler70},
\cite{Stoler74} and our recent article \cite{Acosta-Suslov14} on the
degenerate parametric amplifiers \cite{Raiford70}, \cite{Raiford74}. Moreover,
traditionally, the interaction picture \cite{Mollow67},
\cite{Scully:Zubairy97}, \cite{Walls:Milburn} is commonly used for these
exactly solvable models in quantum nonlinear optics even though the statistics
is postulated in the Schr\"{o}dinger picture \cite{Klyshko94},
\cite{Klyshko98}. This is why, in this article, we continue to study the
statistical properties of output squeezed quanta in terms of explicit
solutions of certain Ermakov-type system \cite{Lan:Lop:Sus} (the initial
data/constants of motion identify the squeezed states under consideration). In
particular, the time-dependent probability amplitudes for a generic model of
variable media are derived. Once again, we elaborate on time development of
the corresponding photon states in Fock's space in the Schr\"{o}dinger
picture. The corresponding evolution operator and the transformation to
Heisenberg's picture are also discussed. In order to achieve these goals, we
utilize a unified approach to generalized harmonic oscillators discussed in
detail in several recent publications \cite{Cor-Sot:Lop:Sua:Sus},
\cite{Cor-Sot:Sus}, \cite{CorSus11}, \cite{KoutSuaSus}, \cite{KretalSus13},
\cite{Lan:Lop:Sus}, \cite{Lop:Sus:VegaGroup}, \cite{SanSusVin} (see also
\cite{Dod:Mal:Man75}, \cite{Dodonov:Man'koFIAN87}, \cite{KrLanSus150},
\cite{Malk:Man:Trif73} for the classical accounts).

Dynamical features of the squeezed $n$-photon configurations, which are not
usually discussed in cavity QED in detail, is our primarily motivation for
writing of this comment. The article is organized as follows. To begin with,
we briefly review the method of dynamical invariants for quantization of
radiation field in order to make our presentation as self-contained as
possible. Solutions of Heisenberg's equation of motion are found with the help
of an evolution operator and then independently verified by a direct
substitution. Finally, probability amplitudes with respect to the number
states are re-derived in an operator form (for a generic model of variable
medium). As a by-product, their time evolution is established, by making use
of a general solution of the Ermakov-type system. Potential applications of
these results to some recent experiments are mentioned in the end.

\section{Quantization of Radiation Field in Variable Media}

In this article, we consider the quantization of radiation field in a variable
dielectric medium in the Schr\"{o}dinger picture, as outlined in
\cite{Ber:Lif:Pit}, \cite{Heitler57} (in vacuum), by using the method of
dynamical invariants originally developed in \cite{Dod:Mal:Man75},
\cite{Dodonov:Man'koFIAN87}, \cite{Malk:Man:Trif73} and revisited in
\cite{Cor-Sot:Sua:SusInv}, \cite{SanSusVin}, \cite{Suslov10}. In summary, we
separate spatial and time variables in the phenomenological Maxwell equations
with factorized electric permittivity and magnetic permeability (tensors)
\cite{Dod:Klim:Nik93} (see also Appendix in \cite{KretalSus13}). Under proper
boundary conditions, solution of the spatial equations provides a complete set
of eigenfunctions and eigenvalues for the corresponding field expansions. In
the time domain, we follow the mathematical technique of the field
quantization for a variable quadratic system in an abstract (Fock-)Hilbert
space recently discussed in \cite{KretalSus13},\ primarily concentrating on a
single mode of the radiation field. In this picture, the electric and magnetic
fields are represented by certain time-independent operators and the time
evolution is governed by the Schr\"{o}dinger equation for the field state
vector $\left\vert \psi\left(  t\right)  \right\rangle :$
\begin{equation}
i\frac{d}{dt}\left\vert \psi\left(  t\right)  \right\rangle =\widehat{H}%
\left(  t\right)  \left\vert \psi\left(  t\right)  \right\rangle
\label{Schroedinger}%
\end{equation}
with a variable quadratic Hamiltonian of the form%
\begin{equation}
\widehat{H}\left(  t\right)  =a(t)\left.  \widehat{p}\right.  ^{2}+b(t)\left.
\widehat{q}\right.  ^{2}+c(t)\left.  \widehat{q}\right.  \left.
\widehat{p}\right.  -id(t)-f(t)\left.  \widehat{q}\right.  -g(t)\left.
\widehat{p}\right.  , \label{Hamiltonian}%
\end{equation}
where $a,$ $b,$ $c,$ $d,$ $f,$ and $g$ are suitable real-valued functions of
time only and the time-independent operators $\widehat{q}$ and $\widehat{p}$
obey the canonical commutation rule $\left[  \left.  \widehat{q}\right.
,\left.  \widehat{p}\right.  \right]  =i$ (in the units of $\hbar).$

The time-dependent annihilation $\widehat{b}(t)$ and creation $\widehat{b}%
^{\dagger}(t)$ operators (linear integrals of motion \cite{Dod:Mal:Man75},
\cite{Malk:Man:Trif73}, \cite{SanSusVin}) are described by Theorem~1 of
\cite{KretalSus13},%
\begin{align}
&  \widehat{b}(t)=\frac{e^{-2i\gamma\left(  t\right)  }}{\sqrt{2}}\left(
\beta\left(  t\right)  \widehat{q}+\varepsilon\left(  t\right)  +i\frac
{\widehat{p}-2\alpha\left(  t\right)  \widehat{q}-\delta\left(  t\right)
}{\beta\left(  t\right)  }\right)  ,\label{aacross(t)QED}\\
&  \widehat{b}^{\dagger}(t)=\frac{e^{2i\gamma\left(  t\right)  }}{\sqrt{2}%
}\left(  \beta\left(  t\right)  \widehat{q}+\varepsilon\left(  t\right)
-i\frac{\widehat{p}-2\alpha\left(  t\right)  \widehat{q}-\delta\left(
t\right)  }{\beta\left(  t\right)  }\right)  ,\nonumber
\end{align}
in terms of solutions of the Ermakov-type system%
\begin{equation}
\frac{d\alpha}{dt}+b+2c\alpha+4a\alpha^{2}=a\beta^{4}, \label{SysA}%
\end{equation}%
\begin{equation}
\frac{d\beta}{dt}+\left(  c+4a\alpha\right)  \beta=0, \label{SysB}%
\end{equation}%
\begin{equation}
\frac{d\gamma}{dt}+a\beta^{2}=0 \label{SysC}%
\end{equation}
and%
\begin{equation}
\frac{d\delta}{dt}+\left(  c+4a\alpha\right)  \delta-f-2g\alpha=2a\beta
^{3}\varepsilon, \label{SysD}%
\end{equation}%
\begin{equation}
\frac{d\varepsilon}{dt}-\left(  g-2a\delta\right)  \beta=0, \label{SysE}%
\end{equation}%
\begin{equation}
\frac{d\kappa}{dt}-g\delta+a\delta^{2}=a\beta^{2}\varepsilon^{2} \label{SysF}%
\end{equation}
subject to a given initial data. These dynamical invariants obey the canonical
commutation rule $\widehat{b}(t)\widehat{b}^{\dagger}(t)-\widehat{b}^{\dagger
}(t)\widehat{b}(t)=1$ at all times. The corresponding evolution
(Heisenberg-type) equations take the form \cite{KretalSus13}:%
\begin{equation}
i\frac{d}{dt}\widehat{b}\left(  t\right)  +\left[  \widehat{b}\left(
t\right)  ,\ \widehat{H}\left(  t\right)  \right]  =0,\qquad i\frac{d}%
{dt}\widehat{b}^{\dagger}\left(  t\right)  +\left[  \widehat{b}^{\dagger
},\ \widehat{H}\left(  t\right)  \right]  =0. \label{EquayionInvariants}%
\end{equation}
Different forms of solutions of the Ermakov-type system have been found in
\cite{KoutSuaSus}, \cite{KretalSus13}, \cite{Lan:Lop:Sus}, and \cite{Lop:Sus:VegaGroup} in
general. Explicit special cases are discussed in \cite{AcostaSuazo13},
\cite{Acosta-Suslov14}, \cite{KoutSuaSus}, \cite{LopSusVegaHarm}, \cite{Marhic78}.

The corresponding dynamical Fock states,%
\begin{equation}
\left\vert \psi_{n}\left(  t\right)  \right\rangle =\frac{1}{\sqrt{n!}}\left(
\widehat{b}^{\dagger}\left(  t\right)  \right)  ^{n}\left\vert \psi_{0}\left(
t\right)  \right\rangle ,\qquad\widehat{b}\left(  t\right)  \left\vert
\psi_{0}\left(  t\right)  \right\rangle =0, \label{DynamicFock}%
\end{equation}
where the phase $\kappa\left(  t\right)  $ finally shows up, can be obtained
in a standard fashion \cite{Ber:Lif:Pit}, \cite{Fock32-2}, \cite{Fock34-3}
(see also \cite{KobManin89}, in particular, dialogues 8 and 9 and section 3.4,
and (\ref{TimeSqueezeOper})). Under a certain condition, they do satisfy the
time-dependent Schr\"{o}dinger equation (\ref{Schroedinger}) (see
\cite{Fock28-2}, Lemma~2 of Ref.~\cite{KretalSus13} and (\ref{TimeSqueezeOper}%
) for more details). The state $\left\vert \psi_{n}\left(  t\right)
\right\rangle $ can be associated with the multi-parameter squeezed $n$-photon
radiation field, whereas the initial data of Ermakov-type system provide the
corresponding classical constants/integrals of motion, and we elaborate on it
in the next section (see also \cite{BB06}, \cite{Heisenberg76a},
\cite{Heisenberg76b}, \cite{Klyshko94}, \cite{Klyshko98}).

\section{The Canonical Transformation and Evolution Operator}

In the case of the standard time-independent annihilation and creation
operators for a given harmonic mode $\omega,$%
\begin{equation}
\widehat{a}=\frac{1}{\sqrt{2\omega}}\left(  \omega\widehat{q}+i\left.
\widehat{p}\right.  \right)  ,\qquad\widehat{a}^{\dagger}=\frac{1}%
{\sqrt{2\omega}}\left(  \omega\widehat{q}-i\left.  \widehat{p}\right.
\right)  ,\qquad\left[  \widehat{a},\ \widehat{a}^{\dagger}\right]  =1,
\label{aaspq}%
\end{equation}
one can formally get with the help of the Baker--Campbell--Hausdorff formula
that%
\begin{equation}
e^{i\left(  \widehat{a}^{\dagger}\ \widehat{a}\right)  \omega t}\left.
\widehat{a}\right.  e^{-i\left(  \widehat{a}^{\dagger}\ \widehat{a}\right)
\omega t}=\left.  \widehat{a}\right.  e^{-i\omega t}. \label{InterIdentity}%
\end{equation}
Two other familiar identities are valid:%
\begin{align}
&  e^{\xi^{\ast}\left.  \widehat{a}\right.  -\xi\left.  \widehat{a}^{\dagger
}\right.  }\left.  \widehat{a}\right.  e^{-\xi^{\ast}\left.  \widehat{a}%
\right.  +\xi\left.  \widehat{a}^{\dagger}\right.  }=\left.  \widehat{a}%
\right.  +\xi,\label{OperIds}\\
&  e^{\left(  e^{2i\varphi}\left.  \widehat{a}\right.  ^{2}-e^{-2i\varphi
}\left(  \widehat{a}^{\dagger}\right)  ^{2}\right)  \tau/2}\left.
\widehat{a}\right.  e^{-\left(  e^{2i\varphi}\left.  \widehat{a}\right.
^{2}-e^{-2i\varphi}\left(  \widehat{a}^{\dagger}\right)  ^{2}\right)  \tau
/2}\label{OpIdSqueeze}\\
&  \qquad=\left(  \cosh\tau\right)  \left.  \widehat{a}\right.  +e^{-2i\varphi
}\left(  \sinh\tau\right)  \left.  \widehat{a}^{\dagger}\right. \nonumber
\end{align}
for the single-mode displacement \cite{Glauber63} and squeeze \cite{Stoler70},
\cite{Stoler71} operators, respectively (see, for example, \cite{Schumaker86},
\cite{SchumakerCaves85} for more details). With the aid of
(\ref{InterIdentity}) and (\ref{OperIds})--(\ref{OpIdSqueeze}), we obtain the
canonical transformation of the form:%
\begin{equation}
\widehat{b}\left(  t\right)  =\boldsymbol{U}\left(  t\right)  \left.
\widehat{a}\right.  \boldsymbol{U}^{-1}\left(  t\right)  =\frac{e^{-2i\gamma}%
}{\sqrt{2}}\left(  \beta\widehat{q}+\varepsilon+i\frac{\widehat{p}%
-2\alpha\widehat{q}-\delta}{\beta}\right)  \label{BCanonical}%
\end{equation}
for the following unitary operator:%
\begin{equation}
\boldsymbol{U}\left(  t\right)  =e^{i\left(  \widehat{a}^{\dagger
}\ \widehat{a}\right)  \theta}e^{\left(  e^{2i\varphi}\left.  \widehat{a}%
\right.  ^{2}-e^{-2i\varphi}\left(  \widehat{a}^{\dagger}\right)  ^{2}\right)
\tau/2}e^{\xi^{\ast}\left.  \widehat{a}\right.  -\xi\left.  \widehat{a}%
^{\dagger}\right.  }e^{2i\left(  \widehat{a}^{\dagger}\ \widehat{a}\right)
\gamma} \label{UofT}%
\end{equation}
provided%
\begin{align}
&  \frac{1}{\sqrt{\omega}}\left(  \beta-\frac{2i\alpha}{\beta}\right)
+\frac{\sqrt{\omega}}{\beta}=2e^{-i\theta}\cosh\tau,\label{ABTwoParameters}\\
&  \frac{1}{\sqrt{\omega}}\left(  \beta-\frac{2i\alpha}{\beta}\right)
-\frac{\sqrt{\omega}}{\beta}=2e^{i\left(  \theta-2\varphi\right)  }\sinh\tau,
\label{ABTParametersPhy}%
\end{align}
and%
\begin{equation}
\xi\sqrt{2}=\varepsilon-i\frac{\delta}{\beta}. \label{CShift}%
\end{equation}
As a result, the time-dependent parameters of our single-mode
\textquotedblleft multi-parameter squeeze/evolution operator\textquotedblright%
\ (\ref{UofT}), namely, $\theta(t),$ $\tau(t),$ $\varphi(t)$ and $\xi(t),$ are
determined in terms of solutions of the corresponding Ermakov-type system as
follows%
\begin{align}
&  \tan\theta\left(  t\right)  =\frac{2\alpha}{\beta^{2}+\omega},\quad
\tan2\varphi\left(  t\right)  =\frac{4\alpha\beta^{2}}{\beta^{4}-4\alpha
^{2}-\omega^{2}},\label{ParametersThetaTau}\\
&  4\left[  \cosh\tau\left(  t\right)  \right]  ^{2}=\left(  \frac{\beta
}{\sqrt{\omega}}+\frac{\sqrt{\omega}}{\beta}\right)  ^{2}+\frac{4\alpha^{2}%
}{\omega\beta^{2}},\label{ParameterTau}\\
&  4\left[  \sinh\tau\left(  t\right)  \right]  ^{2}=\left(  \frac{\beta
}{\sqrt{\omega}}-\frac{\sqrt{\omega}}{\beta}\right)  ^{2}+\frac{4\alpha^{2}%
}{\omega\beta^{2}} \label{ParameterTauSh}%
\end{align}
(see also (\ref{CShift})). To the best of our knowledge, these relations are
omitted in the available literature and, therefore, may be considered as a
main result of this article.

Assuming that the vacuum state is nondegenerate, one gets $\left.
\widehat{a}\right.  \left(  \boldsymbol{U}^{-1}\left(  t\right)  \left\vert
\psi_{0}\left(  t\right)  \right\rangle \right)  =0$ and $\left\vert \psi
_{0}\left(  t\right)  \right\rangle =\boldsymbol{U}\left(  t\right)
\left\vert 0\right\rangle .$ Then,
\begin{equation}
\left\vert \psi_{n}\left(  t\right)  \right\rangle =\frac{1}{\sqrt{n!}}\left(
\widehat{b}^{\dagger}\left(  t\right)  \right)  ^{n}\left\vert \psi_{0}\left(
t\right)  \right\rangle =\boldsymbol{U}\left(  t\right)  \left(  \frac
{1}{\sqrt{n!}}\left(  \widehat{a}^{\dagger}\right)  ^{n}\left\vert
0\right\rangle \right)  =\boldsymbol{U}\left(  t\right)  \left\vert
n\right\rangle \label{TimeSqueezeOper}%
\end{equation}
in terms of standard Fock's number states $\left\vert n\right\rangle .$ With
the help of expansion, $\left\vert \psi\right\rangle _{0}=\sum_{n=0}^{\infty
}c_{n}\ \left\vert n\right\rangle ,$ the action of operator $\boldsymbol{U}%
\left(  t\right)  $ can be extended, by completeness and linearity, to a
certain class of arbitrary initial data,%
\begin{equation}
\boldsymbol{U}\left(  t\right)  \left\vert \psi\right\rangle _{0}=\sum
_{n=0}^{\infty}c_{n}\ \left\vert \psi_{n}\left(  t\right)  \right\rangle
=\left\vert \psi\left(  t\right)  \right\rangle . \label{ArbitraryInitialData}%
\end{equation}
Therefore, our evolution operator $\boldsymbol{U}\left(  t\right)  $
satisfies, formally, the time-dependent Schr\"{o}dinger equation
(\ref{Schroedinger}), whereas $\boldsymbol{U}\left(  0\right)  \left\vert
\psi\right\rangle _{0}=\left\vert \psi\left(  0\right)  \right\rangle .$ Here,
$\boldsymbol{U}\left(  0\right)  \neq\ id,$ not the identity operator in
general, but a composition involving the familiar time-independent
displacement and squeeze operators.

As a result, one may conclude that the minimum-uncertainty squeezed states,
which are important in most applications, occur when $\alpha\left(
t_{\text{min}}\right)  =0$ and $n=0$ (see, for example, Eq.~(5.5) of
Ref.~\cite{KretalSus13}). Indeed, only at these instances, by
(\ref{ParametersThetaTau}) the following conditions hold, $\theta\left(
t_{\text{min}}\right)  =\varphi\left(  t_{\text{min}}\right)  =\alpha\left(
t_{\text{min}}\right)  =0,$ and a traditional definition of single-mode
squeeze operator from Refs.~\cite{Schumaker86}, \cite{SchumakerCaves85},
\cite{Stoler70}, \cite{Stoler71} can be used, say, \textquotedblleft
stroboscopically\textquotedblright\cite{Stoler74}. In general, the evolution
operator in (\ref{UofT}) and (\ref{TimeSqueezeOper}) should be applied for our
initial \textquotedblleft multi-parameter squeezed number
states\textquotedblright\ given by $\left\vert \psi_{n}\left(  0\right)
\right\rangle =\boldsymbol{U}\left(  0\right)  \left\vert n\right\rangle $
(see also \cite{KrySusVega13}, \cite{Lop:Sus:VegaGroup}, \cite{LopSusVegaHarm}%
, \cite{Marhic78}). The traditional squeeze operator corresponds to the
special initial data $\alpha\left(  0\right)  =\theta\left(  0\right)
=\varphi\left(  0\right)  =0$ and $\tau\left(  0\right)  =\left(  1/2\right)
\ln\left(  \beta^{2}\left(  0\right)  /\omega\right)  .$

In summary, our mathematical analysis identifies(/supports a concept of) the
multi-parameter squeezed $n$-photon (polarized) states of radiation field in
the generic model of variable media \cite{Dod:Klim:Nik93}. Their
time-evolution is described in terms of solutions of the Ermakov-type system.

\section{Solving Heisenberg's Equations of Motion}

In this article, we use the Schr\"{o}dinger picture for investigation of the
quantum statistics of the field.\footnote{A detailed analysis of different
representations of quantum mechanics and quantum optics is given in
Refs.~\cite{Klyshko94}, \cite{Klyshko96}, \cite{Klyshko98}.} On the contrary,
in the Heisenberg picture, in view of (\ref{BCanonical}) one gets%
\begin{equation}
\widehat{q}\left(  t\right)  =\boldsymbol{U}^{-1}\left(  t\right)  \left.
\widehat{q}\right.  \boldsymbol{U}\left(  t\right)  =\frac{1}{\beta\sqrt{2}%
}\left(  e^{2i\gamma}\left.  \widehat{a}\right.  +\left.  \widehat{a}%
^{\dagger}\right.  e^{-2i\gamma}\right)  -\frac{\varepsilon}{\beta}
\label{HeisenbergQ}%
\end{equation}
and%
\begin{align}
\widehat{p}\left(  t\right)   &  =\boldsymbol{U}^{-1}\left(  t\right)  \left.
\widehat{p}\right.  \boldsymbol{U}\left(  t\right)  =\frac{e^{2i\gamma}}%
{\beta\sqrt{2}}\left(  2\alpha-i\beta^{2}\right)  \left.  \widehat{a}\right.
\label{HeisenbergP}\\
&  +\frac{e^{-2i\gamma}}{\beta\sqrt{2}}\left(  2\alpha+i\beta^{2}\right)
\left.  \widehat{a}^{\dagger}\right.  +\delta-\frac{2\alpha\varepsilon}{\beta
}.\nonumber
\end{align}
The Heisenberg equations of motion hold,%
\begin{equation}
i\frac{d}{dt}\widehat{p}\left(  t\right)  =\left[  \widehat{p}\left(
t\right)  ,\ \widehat{\mathcal{H}}\left(  t\right)  \right]  ,\qquad i\frac
{d}{dt}\widehat{q}\left(  t\right)  =\left[  \widehat{q}\left(  t\right)
,\ \widehat{\mathcal{H}}\left(  t\right)  \right]  , \label{HeisenbergPQ}%
\end{equation}
where $\ \widehat{\mathcal{H}}\left(  t\right)  =\boldsymbol{U}^{-1}\left(
t\right)  \widehat{H}\left(  t\right)  \boldsymbol{U}\left(  t\right)  .$ This
can also be verified by a direct substitution with the help of a computer
algebra system.\footnote{See complementary \textsl{Mathematica} notebook:
HeisenbergEquations.nb.}

Here, all information about the state of radiation field is encoded into our
time-dependent operators (\ref{HeisenbergQ})--(\ref{HeisenbergP}) in the form
of initial data/constants of motion of the Ermakov-type system (\ref{SysA}%
)--(\ref{SysF}). The corresponding field evolution is completely determined by
explicit solutions of this system subject to given initial data. In
particular, $\widehat{q}\left(  0\right)  =\widehat{q}$ and $\widehat{p}%
\left(  0\right)  =\widehat{p},$ when $\alpha\left(  0\right)  =\gamma\left(
0\right)  =\delta\left(  0\right)  =\varepsilon\left(  0\right)  =0$ and
$\beta\left(  0\right)  =\sqrt{\omega}.$ It is worth noting that our solutions
correspond to the most general set of (known) quantum numbers for the state of
radiation field under consideration. Moreover, Heisenberg's operators are
given by%
\begin{align}
&  \widehat{a}\left(  t\right)  =\frac{1}{\sqrt{2\omega}}\left(
\omega\widehat{q}\left(  t\right)  +i\left.  \widehat{p}\left(  t\right)
\right.  \right)  =\frac{e^{2i\gamma}}{2\beta\sqrt{\omega}}\left(
\omega+\beta^{2}+2i\alpha\right)  \left.  \widehat{a}\right.
\label{a(t)Heisenberg}\\
&  \quad+\frac{e^{-2i\gamma}}{2\beta\sqrt{\omega}}\left(  \omega-\beta
^{2}+2i\alpha\right)  \left.  \widehat{a}^{\dagger}\right.  -\frac{1}%
{\sqrt{2\omega}}\left[  \frac{\omega\varepsilon}{\beta}-i\left(  \delta
-\frac{2\alpha\varepsilon}{\beta}\right)  \right] \nonumber
\end{align}
and%
\begin{align}
&  \widehat{a}^{\dagger}\left(  t\right)  =\frac{1}{\sqrt{2\omega}}\left(
\omega\widehat{q}\left(  t\right)  -i\left.  \widehat{p}\left(  t\right)
\right.  \right)  =\frac{e^{2i\gamma}}{2\beta\sqrt{\omega}}\left(
\omega-\beta^{2}-2i\alpha\right)  \left.  \widehat{a}\right.
\label{across(t)Heisenberg}\\
&  \quad+\frac{e^{-2i\gamma}}{2\beta\sqrt{\omega}}\left(  \omega+\beta
^{2}-2i\alpha\right)  \left.  \widehat{a}^{\dagger}\right.  -\frac{1}%
{\sqrt{2\omega}}\left[  \frac{\omega\varepsilon}{\beta}+i\left(  \delta
-\frac{2\alpha\varepsilon}{\beta}\right)  \right]  ,\nonumber
\end{align}
once again, in terms of solutions of the Ermakov-type system, which can be
derived from (\ref{HeisenbergPQ}) (see \textsl{Mathematica} notebook:
HeisenbergEquations.nb). Thus%
\begin{align}
\left[  \widehat{a}\left(  t\right)  ,\ \widehat{a}\left(  t^{\prime}\right)
\right]   &  =A\left(  t\right)  B\left(  t^{\prime}\right)  -A\left(
t^{\prime}\right)  B\left(  t\right)  ,\label{AAT}\\
\left[  \widehat{a}\left(  t\right)  ,\ \widehat{a}^{\dagger}\left(
t^{\prime}\right)  \right]   &  =A\left(  t\right)  A^{\ast}\left(  t^{\prime
}\right)  -B\left(  t\right)  B^{\ast}\left(  t^{\prime}\right)
,\label{AACT}\\
\left[  \widehat{a}^{\dagger}\left(  t\right)  ,\ \widehat{a}^{\dagger}\left(
t^{\prime}\right)  \right]   &  =B^{\ast}\left(  t\right)  A^{\ast}\left(
t^{\prime}\right)  -A^{\ast}\left(  t\right)  B^{\ast}\left(  t^{\prime
}\right)  . \label{ACACT}%
\end{align}
Here, the asterisk represents the complex conjugate and, by definition,%
\begin{equation}
A\left(  t\right)  =\frac{e^{2i\gamma}}{2\beta\sqrt{\omega}}\left(
\omega+\beta^{2}+2i\alpha\right)  ,\qquad B\left(  t\right)  =\frac
{e^{-2i\gamma}}{2\beta\sqrt{\omega}}\left(  \omega-\beta^{2}+2i\alpha\right)
. \label{ABT}%
\end{equation}
These relations reveal the fact of noncommutativity of the radiation field
operators at different times.

The expectation values and variances of Heisenberg's operators $\left.
\widehat{q}\left(  t\right)  \right.  $ and $\left.  \widehat{p}\left(
t\right)  \right.  $ with respect to the standard Fock number states
$\left\vert n\right\rangle $ coincide, of course, with those found in
Ref.~\cite{KretalSus13}, section~5, in the Schr\"{o}dinger picture.

\noindent {\it{Example.}}
Special cases include the degenerate parametric amplifiers
\cite{AcostaSuazo13}, \cite{Acosta-Suslov14}, \cite{Mollow73},
\cite{Raiford70}, \cite{Raiford74}, \cite{Stoler74}, which provide a natural
mechanism of creation of the multi-parameter squeezed states of light (see
also \cite{KrySusVega13}, \cite{Marhic78} and the references therein). For the
Hamiltonian of the form%
\begin{equation}
\widehat{H}\left(  t\right)  =\frac{\omega}{2}\left(  \widehat{a}%
\ \widehat{a}^{\dagger}+\widehat{a}^{\dagger}\ \widehat{a}\right)
+\frac{\lambda}{2i}\left(  e^{2i\omega t}\ \left.  \widehat{a}\right.
^{2}-e^{-2i\omega t}\left(  \widehat{a}^{\dagger}\right)  ^{2}\right)  ,
\label{DPAII}%
\end{equation}
our solutions of Heisenberg's equations (\ref{HeisenbergPQ}) are given by%
\begin{align}
\ \widehat{a}\left(  t\right)   &  =\boldsymbol{U}^{-1}\left(  t\right)
\left.  \widehat{a}\right.  \boldsymbol{U}\left(  t\right)  =\frac{\omega
e^{2\lambda t}+\beta^{2}\left(  0\right)  +2i\alpha\left(  0\right)  }%
{2\beta\left(  0\right)  \sqrt{\omega}}e^{-\left(  \lambda+i\omega\right)
t}\left.  \widehat{a}\right. \label{a(t)DPAII}\\
&  +\frac{\omega e^{2\lambda t}-\beta^{2}\left(  0\right)  +2i\alpha\left(
0\right)  }{2\beta\left(  0\right)  \sqrt{\omega}}e^{-\left(  \lambda
+i\omega\right)  t}\left.  \widehat{a}^{\dagger}\right. \nonumber\\
&  +i\frac{\beta\left(  0\right)  \delta\left(  0\right)  -2\alpha\left(
0\right)  \varepsilon\left(  0\right)  +i\omega e^{2\lambda t}\varepsilon
\left(  0\right)  }{\beta\left(  0\right)  \sqrt{2\omega}}e^{-\left(
\lambda+i\omega\right)  t}\nonumber
\end{align}
\newline and%
\begin{align}
\ \widehat{a}^{\dagger}\left(  t\right)   &  =\boldsymbol{U}^{-1}\left(
t\right)  \left.  \widehat{a}^{\dagger}\right.  \boldsymbol{U}\left(
t\right)  =\frac{\omega e^{2\lambda t}-\beta^{2}\left(  0\right)
-2i\alpha\left(  0\right)  }{2\beta\left(  0\right)  \sqrt{\omega}}e^{\left(
i\omega-\lambda\right)  t}\left.  \widehat{a}\right. \label{across(t)DPAII}\\
&  +\frac{\omega e^{2\lambda t}+\beta^{2}\left(  0\right)  -2i\alpha\left(
0\right)  }{2\beta\left(  0\right)  \sqrt{\omega}}e^{\left(  i\omega
-\lambda\right)  t}\left.  \widehat{a}^{\dagger}\right. \nonumber\\
&  +i\frac{2\alpha\left(  0\right)  \varepsilon\left(  0\right)  -\beta\left(
0\right)  \delta\left(  0\right)  +i\omega e^{2\lambda t}\varepsilon\left(
0\right)  }{\beta\left(  0\right)  \sqrt{2\omega}}e^{\left(  i\omega
-\lambda\right)  t}.\nonumber
\end{align}
The particular solutions, corresponding to the initial data $\alpha\left(
0\right)  =\gamma\left(  0\right)  =\delta\left(  0\right)  =\varepsilon
\left(  0\right)  =0$ and $\beta\left(  0\right)  =\sqrt{\omega},$ are
originally found in \cite{Mollow73}.

\section{On Variable Photon Statistics}

The explicit form of the unitary operator (\ref{UofT}) allows us to
(re-)evaluate the time-dependent photon amplitudes with respect to the Fock
basis, namely,%
\begin{equation}
\left\vert \psi_{n}\left(  t\right)  \right\rangle =\sum_{m=0}^{\infty}\left(
e^{2i\gamma}\left[  \left(  \sum_{k=0}^{\infty}S_{mk}D_{kn}\right)  \right]
e^{im\theta}\right)  \left\vert m\right\rangle , \label{Amplitudes}%
\end{equation}
in a pure algebraic fashion similar to Refs.~\cite{Glauber63}, \cite{Marian91}%
, \cite{Marian92}, \cite{VenkSatya85} (but for the generic variable quadratic
Hamiltonian). Here, the matrix elements of the displacement operator are given
by%
\begin{align}
&  D_{mn}\left(  \xi\right)  =\left\langle m\left\vert e^{\xi^{\ast}\left.
\widehat{a}\right.  -\xi\left.  \widehat{a}^{\dagger}\right.  }\right\vert
n\right\rangle \label{DisplacementMatrix}\\
&  \quad=e^{-\left\vert \xi\right\vert ^{2}/2}\frac{\left(  -\xi\right)
^{m}\left(  \xi^{\ast}\right)  ^{n}}{\sqrt{m!n!}}\ _{2}F_{0}\left(
-n,-m;-\frac{1}{\left\vert \xi\right\vert ^{2}}\right)  ,\nonumber
\end{align}
where parameter $\xi$ is determined by (\ref{CShift}) in terms of solutions of
Ermakov-type system. (Familiar relations with the Charlier polynomials are
discussed in \cite{Ni:Su:Uv}.)

In a similar fashion, the matrix elements of the squeeze operator can be
readily evaluated. By using an important factorization identity
\cite{Schumaker86}, \cite{SchumakerCaves85},%
\begin{align}
e^{\left(  e^{2i\varphi}\left.  \widehat{a}\right.  ^{2}-e^{-2i\varphi}\left(
\widehat{a}^{\dagger}\right)  ^{2}\right)  \tau/2}  &  =e^{-\left(
1/2\right)  e^{-2i\varphi}\tanh\tau\left(  \left.  \widehat{a}^{\dagger
}\right.  \right)  ^{2}}\label{Identity}\\
&  \times e^{-\ln\cosh\tau\left(  \widehat{a}^{\dagger}\left.  \widehat{a}%
\right.  +1/2\right)  }e^{\left(  1/2\right)  e^{2i\varphi}\tanh\tau\left(
\left.  \widehat{a}\right.  \right)  ^{2}},\nonumber
\end{align}
one can derive the following expression:%
\begin{align}
&  \medskip S_{mn}\left(  \alpha,\beta\right)  =\left\langle m\left\vert
e^{\left(  e^{2i\varphi}\left.  \widehat{a}\right.  ^{2}-e^{-2i\varphi}\left(
\widehat{a}^{\dagger}\right)  ^{2}\right)  \tau/2}\right\vert n\right\rangle
\label{SqueezeOperator}\\
&  =\sqrt{\frac{m!n!\pi}{2^{m+n}\cosh\tau}}\dfrac{\left(  -e^{-2i\varphi}%
\sinh\tau\right)  ^{\left(  m-n\right)  /2}\left(  \cosh\tau\right)
^{-\left(  m+n\right)  /2}}{\Gamma\left(  \frac{m-n}{2}+1\right)
\Gamma\left(  \frac{n}{2}+1\right)  \Gamma\left(  \frac{n+1}{2}+1\right)
}\nonumber\\
&  \qquad\times\ _{2}F_{1}\left(
\begin{array}
[c]{c}%
(1-n)/2,\quad-n/2\medskip\\
1+\left(  m-n\right)  /2
\end{array}
;\quad-\sinh^{2}\tau\right)  ,\qquad m\geq n.\nonumber
\end{align}
With the help of familiar transformations of the terminating hypergeometric
functions \cite{An:Ask:Roy}, we finally obtain the non-vanishing matrix
elements as follows%
\begin{align}
\medskip S_{mn}\left(  \alpha,\beta\right)   &  =\dfrac{\left(  -1\right)
^{m/2}e^{-i\left(  m-n\right)  \varphi}}{\sqrt{\cosh\tau}}\left[
\dfrac{\left(  1/2\right)  _{m/2}\left(  1/2\right)  _{n/2}}{\left(
m/2\right)  !\left(  n/2\right)  !}\right]  ^{1/2}\label{SMartixEven}\\
&  \times\left(  \tanh\tau\right)  ^{\left(  m+n\right)  /2}\ _{2}F_{1}\left(
\begin{array}
[c]{c}%
-n/2,\quad-m/2\medskip\\
1/2
\end{array}
;\quad-\dfrac{1}{\sinh^{2}\tau}\right)  ,\nonumber
\end{align}
if $m,\ n$ are even and%
\begin{align}
\medskip S_{mn}\left(  \alpha,\beta\right)   &  =2\dfrac{\left(  -1\right)
^{(m-1)/2}e^{-i\left(  m-n\right)  \varphi}}{\sinh\tau\sqrt{\cosh\tau}}\left[
\dfrac{\left(  3/2\right)  _{\frac{m-1}{2}}\left(  3/2\right)  _{\frac{n-1}%
{2}}}{\left(  \frac{m-1}{2}\right)  !\left(  \frac{n-1}{2}\right)  !}\right]
^{1/2}\label{SMatrixOdd}\\
&  \times\left(  \tanh\tau\right)  ^{\left(  m+n\right)  /2}\ _{2}F_{1}\left(
\begin{array}
[c]{c}%
(1-n)/2,\quad(1-m)/2\medskip\\
3/2
\end{array}
;\quad-\dfrac{1}{\sinh^{2}\tau}\right)  ,\nonumber
\end{align}
if $m,\ n$ are odd. In view of (\ref{ParametersThetaTau}%
)--(\ref{ParameterTauSh}), the time evolution of these matrix elements is also
found in terms of solutions of the corresponding Ermakov-type system for the
generic model of a variable medium. (The latter hypergeometric functions are
related to Meixner's polynomials \cite{Ni:Su:Uv}.)

In some cases, the complex parametrization and quasi-invariants of the
Ermakov-type system \cite{KretalSus13} may help to simplify the arguments of
the hypergeometric functions. Examples are discussed in \cite{Acosta-Suslov14}
and \cite{KrySusVega13}.

\section{A Conclusion}

Light propagation in a variable medium, say in the process of degenerate
parametric amplification \cite{Acosta-Suslov14}, \cite{Raiford70},
\cite{Raiford74}, \cite{Mollow67}, \cite{Stoler74} in a nonlinear crystal, has
became one of the standard ways of creation of the squeezed states of
radiation field \cite{Breit:Schill:Mlyn97}, \cite{LvRay09}. The latter are
expected to be utilized for detection of gravitational waves \cite{Abadetal11}%
, \cite{BachorRalph04}, \cite{Caves81}, \cite{Demkowiczetal13},
\cite{Hollenhorst79}. Nowadays, advanced experimental techniques allow one to
measure photon correlation functions of input microwave signals
\cite{Bozyigitetal10}, to perform quantum tomography on itinerant microwave
photons \cite{Eichleretal11a}, and to study squeezing of microwave fields
\cite{Eichleretal11b}, \cite{Malletetal11} (see also \cite{Braggioetal13},
\cite{Galeazzietal13} for experimental study of a single-mode thermal field
using a microwave parametric amplifier). Connections with the experimentally
observed dynamical Casimir effect \cite{Lahetal11}, \cite{Wilsonetal11} are
discussed in \cite{DezaelLambrecht10}, \cite{Dod09},
\cite{FujiietalZeilinger11}, \cite{Johanetal10}, \cite{Nationetal12}.
Moreover, nonclassical states of phonons are recently reviewed in
\cite{MisUPN13}. In summary, coherent and, possibly, squeezed and entangled
states of a crystal lattice can be created by using ultrashort laser pulses.
In this article, we have discussed the time-evolution of photon (or phonon)
statistics for the generic model of variable media which may be important in
some of the aforementioned applications.

\noindent\textbf{Acknowledgments.\/} We would like to thank Albert Boggess,
Geza Giedke, and Vladimir~I.~Man'ko for help and encouragement. This research
was partially supported by AFOSR grant FA9550-11-1-0220. One of the authors
(ES) was also supported by the Simons Foundation Grant \# 316295 and by NSF
grants DMS~\#~1535833 and DMS~\#~1151618.


\begin{thebibliography}{99}                                                                                               %


\bibitem {Abadetal11}J.~Abadie et al.,\emph{\ A gravitational wave observatory
operating beyond the quantum shot-noise limit\/}, Nature Physics \textbf{7}
(2011), 962--965.

\bibitem {AcostaSuazo13}P.~B. Acosta-Hum\'{a}nez and E.~Suazo,
\emph{Liouvillian propagators, Riccati equation and differential Galois
theory\/}, J. Phys. A: Math. Theor. \textbf{46} (2013)~\#~45, 455203 (17 pages).

\bibitem {Acosta-Suslov14}P.~B. Acosta-Hum\'{a}nez, S.~I.~Kryuchkov, E.~Suazo,
and S.~K.~Suslov, \emph{Degenerate parametric amplification of squeezed
photons: explicit solutions, statistics, means and variances\/}, J. Nonlinear
Opt. Phys. Mater. \textbf{24} (2015)~\#~2, 1550021 (27 pages).

\bibitem {Agarwal87}G.~S.~Agarwal, \emph{Wigner-function description of
quantum noise in interferometers\/}, J. Mod. Opt. \textbf{34} (1987) \#~6--7, 909--921.

\bibitem {AgarwalAdam88}G.~S.~Agarwal and G. Adam, \emph{Photon-number
distributions for quantum fields generated in nonlinear optical processes\/},
Phys. Rev. A \textbf{38} (1988)~\#~2, 750--752.

\bibitem {An:Ask:Roy}G.~E.~Andrews, R.~Askey, and R.~Roy, \textsl{Special
Functions\/}, Cambridge University Press, Cambridge, 1999.

\bibitem {BachorRalph04}H.-A.~Bachor and T.~C.~Ralph, \textsl{A Guide to
Experiments in Quantum Optics\/}, 2nd ed., Wiley--Vch Publishing Company,
Weinheim, 2004.

\bibitem {Ber:Lif:Pit}V.~B.~Berestetskii, E.~M.~Lifshitz, and
L.~P.~Pitaevskii, \textsl{Relativistic Quantum Theory\/}, Pergamon Press,
Oxford, 1971.

\bibitem {BB06}I.~Bialynicki-Birula, \emph{Photon as a quantum particle\/},
Acta Physica Polonica B \textbf{37} (2006) \#~3, 935--946.

\bibitem {Bialynicka-Birula87}Z.~Bialynicka-Birula and I.~Bialynicki-Birula,
\emph{Space-time description of squeezing\/}, J. Opt. Soc. Am. B \textbf{4}
(1987) \#~10, 1621--1626.

\bibitem {Bozyigitetal10}D.~Bozyigit, C.~Lang, L.~Steffen, J.~M.~Fink,
C.~Eichler, M.~Baur, R.~Bianchetti, P.~J.~Leek, S.~Filipp, M.~P.~da~Silva,
A.~Blais, and A.~Wallraff, \emph{Antibunching of microwave-frequency photons
observed in correlation measurements using linear detectors\/}, Nature Physics
\textbf{7} (2011), February, 154--158.

\bibitem {Braggioetal05}C.~Braggio, G.~Bresso, G.~Carugno, C.~Del Noce,
G.~Galeazzi, A.~Lombardi, A.~Palmieri, G.~Ruoso, and D.~Zanello, \emph{A novel
experimental approach for the detection of the dynamical Casimir effect\/},
Europhys. Lett. \textbf{70} (2005)~\#~6, 754--760.

\bibitem {Braggioetal13}C.~Braggio, G.~Carugno, F.~Della Valle, G.~Galeazzi,
A.~Lombardi, G.~Ruoso, and D.~Zanello, \emph{The measurement of a single-mode
thermal field with a microwave cavity parametric amplifier\/}, New J. Phys.
\textbf{15} (2013), 013044 (9 pages).

\bibitem {Breit:Schill:Mlyn97}G.~Breitenbach, S.~Schiller, and J.~Mlynek,
\emph{Measurement of the quantum states of squeezed light\/},~Nature
\textbf{387} (1997)~May 29, 471--475.

\bibitem {Caves81}C.~M.~Caves, \emph{Quantum-mechatical noise in an
interferometer\/}, Phys. Rev. D \textbf{23} (1981)~\#~8, 1993--1708.

\bibitem {Caves82}C.~M.~Caves, \emph{Quantum limits on noise in linear
amplifiers\/}, Phys. Rev. D \textbf{26} (1982)~\#~8, 1817--1839.

\bibitem {Cor-Sot:Lop:Sua:Sus}R.~Cordero-Soto, R.~M.~L\'{o}pez, E.~Suazo, and
S.~K.~Suslov, \emph{Propagator of a charged particle with a spin in uniform
magnetic and perpendicular electric fields\/}, Lett.~Math.~Phys. \textbf{84}
(2008)~\#~2--3, 159--178.

\bibitem {Cor-Sot:Sua:SusInv}R.~Cordero-Soto, E.~Suazo and S.~K.~Suslov,
\emph{Quantum integrals of motion for variable quadratic Hamiltonians\/}, Ann.
Phys. \textbf{325} (2010)~\#~9, 1884--1912.

\bibitem {Cor-Sot:Sus}R.~Cordero-Soto and S.~K.~Suslov, \emph{Time reversal
for modified oscillators\/}, Theoretical and Mathematical Physics \textbf{162}
(2010)~\#~3, 286--316.

\bibitem {CorSus11}R.~Cordero-Soto and S.~K.~Suslov, \emph{The degenerate
parametric oscillator and Ince's equation\/}, J. Phys. A: Math. Theor.
\textbf{44} (2011)~\#~1, 015101 (9 pages).

\bibitem {Demkowiczetal13}R.~Demkowicz-Dobrza\'{n}ski, K.~Banaszek, and
R.~Schnabel, \emph{Fundamental quantum interferometry bound for the
squeezed-light-enhanced gravitational wave detector GEO 600\/}, Phys. Rev.~A
\textbf{88} (2013) \#~4, 041802(R) (5 pages).

\bibitem {DezaelLambrecht10}F.~X.~Dezael and A.~Lambrecht, \emph{Analogue
Casimir radiation using an optical parametric oscillator\/}, Euro. Phys. Lett.
\textbf{89} (2010), 14001 (5~pages).

\bibitem {Dod09}V.~V.~Dodonov, \emph{Photon distribution in the dynamical
Casimir effect with an account of dissipation\/}, Phys. Rev.~A \textbf{80}
(2009) \#~2, 023814 (11 pages).

\bibitem {Dod:Klim:Nik93}V.~V.~Dodonov, A.~B.~Klimov, and D.~E.~Nikonov,
\emph{Quantum phenomena in nonstationary media\/}, Phys. Rev.~A \textbf{47}
(1993)~\#~5, 4422--4429.

\bibitem {Dod:Mal:Man75}V.~V.~Dodonov, I.~A.~Malkin and V.~I.~Man'ko,
\emph{Integrals of motion, Green functions, and coherent states of dynamical
systems\/}, Int.~J.~Theor.~Phys. \textbf{14} (1975)~\#~1, 37--54.

\bibitem {Dod:Man79}V.~V.~Dodonov and V.~I.~Man'ko, \emph{Coherent states and
the resonance of a quantum damped oscillator\/}, Phys.~Rev.~A \textbf{20}
(1979)~\#~2, 550--560.

\bibitem {Dodonov:Man'koFIAN87}V.~V.~Dodonov and V.~I.~Man'ko,
\emph{Invariants and correlated states of nonstationary quantum systems}, in:
\textsl{Invariants and the Evolution of Nonstationary Quantum Systems\/},
Proceedings of Lebedev Physics Institute, vol. 183, pp. 71-181, Nauka, Moscow,
1987 [in Russian]; English translation published by Nova Science, Commack, New
York, 1989, pp. 103-261.

\bibitem {Dod:Man:Man94}V.~V.~Dodonov, O.~V.~Man'ko, and V.~I.~Man'ko,
\emph{Photon distribution for one-mode mixed light with a generic gaussian
Wigner function}, Phys. Rev. A \textbf{49} (1994), 2993--3001.

\bibitem {Drummomd90}P.~D.~Drummond, \emph{Electromagnetic quantization in
dispersive inhomogeneous nonlinear dielectrics\/},~Phys. Rev. A~\textbf{42}
(1990)~\#~11, 6845--6857.

\bibitem {Eichleretal11a}C.~Eichler, D.~Bozyigit, C.~Lang, L.~Steffen,
J.~Fink, and A.~Wallraff, \emph{Experimental state tomography of itinerant
single microwave photons\/}, Phys. Rev.~Lett. \textbf{106} (2011) \#~22,
220503 (4 pages).

\bibitem {Eichleretal11b}C.~Eichler, D.~Bozyigit, C.~Lang, M.~Baur,
L.~Steffen, J.~Fink, S.~Filipp, and A.~Wallraff, \emph{Observation of two-mode
squeezing in the microwave frequency domain\/}, Phys. Rev.~Lett. \textbf{107}
(2011) \#~11, 113601 (5 pages).

\bibitem {Fock28-2}V.~Fock, \emph{On the relation between the integrals of the
quantum mechanical equations of motion and the Schr\"{o}dinger wave
equation\/}, Zs. Phys. \textbf{49} (1928)~\#~5--6, 323--338; reprinted in:
V.~A.~Fock, \textsl{Selected Works: Quantum Mechanics and Quantum Field
Theory\/}, (L.~D.~Faddeev, L.~A.~Khalfin, and I.~V.~Komarov, Eds.), Chapman \&
Hall/CRC, Boca Raton, London, New York, Washington, D.~C., 2004, pp.~33--49.

\bibitem {Fock32-2}V.~Fock, \emph{Configuration space and second
quantization\/}, Zs. Phys. \textbf{75} (1932)~\#~9--10, 622--647; reprinted
in: V.~A.~Fock, \textsl{Selected Works: Quantum Mechanics and Quantum Field
Theory\/}, (L.~D.~Faddeev, L.~A.~Khalfin, and I.~V.~Komarov, Eds.), Chapman \&
Hall/CRC, Boca Raton, London, New York, Washington, D.~C., 2004, pp.~191--220.

\bibitem {Fock34-3}V.~Fock, \emph{On quantum electrodynamics\/}, Phys. Zs.
Sowjetunion \textbf{6} (1934), 425--469; reprinted in: V.~A.~Fock,
\textsl{Selected Works: Quantum Mechanics and Quantum Field Theory\/},
(L.~D.~Faddeev, L.~A.~Khalfin, and I.~V.~Komarov, Eds.), Chapman \& Hall/CRC,
Boca Raton, London, New York, Washington, D.~C., 2004, pp.~331--368.

\bibitem {FujiietalZeilinger11}T.~Fujii, S.~Matsuo, N.~Hatakenaka,
S.~Kurihara, and A.~Zeilinger, \emph{Quantum circuit analog of the dynamical
Casimir effect\/}, Phys. Rev. B \textbf{84} (2011)~\#~17, 174521 (9 pages).

\bibitem {Galeazzietal13}G.~Galeazzi, A.~Lombardi, G.~Ruoso, C.~Braggio,
G.~Carugno, F.~Della Valle, D.~Zanello, and V.~V.~Dodonov, \emph{Experimental
study of microwave photon statistics under parametric amplification from a
single-mode thermal state in a cavity\/}, Phys. Rev. A \textbf{88}
(2013)~\#~5, 053806 (7 pages).

\bibitem {Glauber63}R.~J.~Glauber, \emph{Coherent and incoherent states of the
radiation field\/}, Phys. Rev. \textbf{131} (1963)~\#~6, 2766--2788.

\bibitem {GlauberNobel}R.~J.~Glauber, \emph{One hundred years of light quanta
(Nobel Lecture)\/}, ChemPhysChem. \textbf{7} (2006)~\#~8, 1618--1639.

\bibitem {GlauberCollect}R.~J.~Glauber, \textsl{Quantum Theory of Optical
Coherence: Selected Papers and Lectures\/}, WILEY-VCH Verlag GmbH \& Co. KGaA,
Weinheim, 2007.

\bibitem {Glauber91}R.~J.~Glauber and M.~Lewenstein, \emph{Quantum optics of
dielectric media\/}, Phys. Rev.~A \textbf{43} (1991)~\#~1, 467--491.

\bibitem {Heisenberg76a}W.~Heisenberg, \emph{The nature of elementary
particles\/}, Phys. Today \textbf{29} (1976)~\#~3, 32--39.

\bibitem {Heisenberg76b}W.~Heisenberg, \emph{Cosmic radiation and fundamental
problems in physics\/}, Naturwissenschaften \textbf{63} (1976)~\#~2, 63--67.

\bibitem {Heitler57}W.~Heitler, \textsl{The Quantum Theory of Radiation\/},
The Clarendon Press, Oxford, 1954.

\bibitem {Hillery09}M.~Hillery, \emph{An introduction to\ the quantum theory
of nonlinear optics\/}, Acta Physica Slovaca \textbf{59} (2009)~\#~1, 1--80.

\bibitem {Hillery84}M.~Hillery and L.~D.~Mlodinow, \emph{Quantization of
electrodynamics in nonlinear dielectric media\/}, Phys. Rev.~A \textbf{30}
(1984)~\#~4, 1860--1865.

\bibitem {Hollenhorst79}J.~N.~Hollenhorst, \emph{Quantum limits on
resonant-mass gravitational-radiation detectors\/}, Phys. Rev. D (1979)~\#~6, 1669--1679.

\bibitem {Johanetal10}J.~R.~Johansson, G.~Johansson, C.~M.~Wilson, and
F.~Nori, \emph{Dynamical Casimir in superconducting microwave circuits\/},
Phys. Rev.~A \textbf{82} (2010) \#~5, 052509 (17 pages).

\bibitem {KimOlivKnight89}M.~S.~Kim, F.~A.~M.~de Oliveira, and P.~L.~Knight,
\emph{Properties of squeezed number states and squeezed thermal states\/},
Phys. Rev.~A \textbf{40} (1989) \#~5, 2494--2503.

\bibitem {Klyshko88}D.~N.~Klyshko, \textsl{Photons and Nonlinear Optics\/},
Gordon and Breach, New York, 1988.

\bibitem {Klyshko94}D.~N.~Klyshko, \emph{Quantum optics: quantum, classical,
and metaphysical aspects\/}, Phys.-Usp. \textbf{37} (1994) \#~11, 1097--1123;
see also: \textit{Annals of the New York Academy of Sciences\/} vol.~755
\textit{Fundamental Problems in Quantum Theory\/} April~1995 pp~13--27.

\bibitem {Klyshko96}D.~N.~Klyshko, \emph{The nonclassical light\/}, Phys.-Usp.
\textbf{39} (1996) \#~6, 573--596.

\bibitem {Klyshko98}D.~N.~Klyshko, \emph{Basic quantum mechanical concepts
from the operational viewpoint\/}, Phys.-Usp. \textbf{41} (1998) \#~9, 885--922.

\bibitem {Klyshko11}D.~N.~Klyshko, \textsl{Physical Foundations of Quantum
Electronics\/}, World Scientific, New Jersey, 2011.

\bibitem {KobManin89}I.~Yu.~Kobzarev and Yu.~I.~Manin, \textsl{Elementary
Particles: Mathematics, Physics and Philosophy\/}, Kluwer, Dordrecht, Boston,
London, 1989.

\bibitem {KolesManko08}A.~A.~Kolesnikov and V.~I.~Man'ko, \emph{Uncertainty
relation and photon-number distribution for one- and two-mode squeezed
light\/}, Journal of Russian Laser Research \textbf{29} (2008)~\#~2, 142--166.

\bibitem {KoutSuaSus}C. Koutschan, E.~Suazo, and S.~K.~Suslov,
\emph{Fundamental laser modes in paraxial optics: from computer algebra and
simulations to experimental observation\/}, Appl. Phys.~B \textbf{121} (2015)~\#~3, 315--336.

\bibitem {KretalSus13}C. Krattenthaler, S.~I.~Kryuchkov, A.~Mahalov, and
S.~K.~Suslov, \emph{On the problem of electromagnetic-field quantization\/},
Int. J. Theor. Phys. \textbf{52} (2013)~\#~12, 4445--4460.

\bibitem {KrLanSus150}S.~I.~Kryuchkov, N.~Lanfear, and S.~K.~Suslov, \emph{The
Pauli-Luba\'{n}ski vector, complex electrodynamics, and photon helicity\/},
Phys. Scr. \textbf{90} (2015)~\#~7, 074065 (8 pages)

\bibitem {KrySusVega13}S.~I.~Kryuchkov, S.~K.~Suslov, and
J.~M.~Vega-Guzm\'{a}n, \emph{The minimum-uncertainty squeezed states for atoms
and photons in a cavity\/}, J. Phys. B: At. Mol. Opt. Phys. \textbf{46}
(2013)~\#~10, 104007 (15 pages), IOP Select and Highlight of 2013.

\bibitem {Lahetal11}P.~L\"{a}hteenm\"{a}ki, G.~S.~Paraoanu, J.~Hassel, and
P.~J.~Hakonen, \emph{Dynamical Casimir effect in a Josephson metamaterial\/},
Proc. Nat. Acad. Sci. \textbf{14} (2013)~\#~11, 4234--4238.

\bibitem {Lan:Lop:Sus}N.~Lanfear, R.~M.~L\'{o}pez, and S.~K.~Suslov,
\emph{Exact wave functions for generalized harmonic oscillators\/}, Journal of
Russian Laser Research \textbf{32} (2011)~\#~4, 352--361.

\bibitem {LeonardPaul95}U.~Leonhard and H.~Paul, \emph{Measuring the quantum
state of light\/}, Prog. Quant. Electr. \textbf{19} (1995), 89--130.

\bibitem {Lop:Sus:VegaGroup}R.~M.~L\'{o}pez, S.~K.~Suslov, and
J.~M.~Vega-Guzm\'{a}n, \emph{Reconstructing the Schr\"{o}dinger groups\/},
Physica Scripta \textbf{87}\ (2013)~\#~3, 038112 (6 pages).

\bibitem {LopSusVegaHarm}R.~M.~L\'{o}pez, S.~K.~Suslov, and
J.~M.~Vega-Guzm\'{a}n, \emph{On a hidden symmetry of quantum harmonic
oscillators\/}, Journal of Difference Equations and Applications, \textbf{19}
(2013)~\#~4, 543--554.

\bibitem {LvRay09}A.~I.~Lvovsky and M.~G.~Raymer, \emph{Continuous-variable
optical quantum-state tomography\/}, Rev. Mod. Phys. \textbf{81} (2009),
January--March, 299--332.

\bibitem {Mah:Sua:Sus13}A.~Mahalov, E.~Suazo, and S.~K.~Suslov, \emph{Spiral
laser beams in inhomogeneous media\/}, Opt. Lett. \textbf{38} (2013)~\#~15, 2763--2766.

\bibitem {Malk:Man:Trif73}I.~A.~Malkin, V.~I.~Man'ko, and D.~A.~Trifonov,
\emph{Linear adiabatic invariants and coherent states\/}, J. Math. Phys.
\textbf{14} (1973) \#~5, 576--582.

\bibitem {Malletetal11}F.~Mallet, M.~A.~Castellanos-Beltran, H.~S.~Ku,
S.~Glancy, E.~Knill, K.~D.~Irwin, G.~C.~Hilton, \emph{Quantum state tomography
of an itinerant squeezed microwave field\/}, Phys. Rev.~Lett. \textbf{106}
(2011) \#~22, 220502 (4 pages).

\bibitem {MankoWuensche97}V.~I.~Man'ko and A.~W\"{u}nsche, \emph{Properties of
squeezed-state excitations\/}, Quantum Semiclass. Opt. \textbf{9} (1997)~\#~3, 381--409.

\bibitem {Marhic78}M.~E.~Marhic, \emph{Oscillating Hermite--Gaussian wave
functions of the harmonic oscillator\/}, Lett. Nuovo Cim. \textbf{22}
(1978)~\#~8, 376--378.

\bibitem {MarhicKumar90}M.~E.~Marhic and P.~Kumar, \emph{Squeezed states with
a thermal photon distribution\/}, Optics Communications \textbf{76}
(1990)~\#~2, 143--146.

\bibitem {Marian91}P.~Marian, \emph{Higher-order squeezing properties and
correlation functions for squeezed number states\/}, Phys. Rev. A \textbf{44}
(1991)~\#~5, 3325--3330.

\bibitem {Marian92}P.~Marian, \emph{Higher-order squeezing and photon
statistics for squeezed thermal states\/}, Phys. Rev. A \textbf{45}
(1992)~\#~3, 2044--2051.

\bibitem {MarianMarianI93}P.~Marian and T.~A.~Marian, \emph{Squeezed states
with thermal noise. I. Photon-number statistics\/}, Phys. Rev. A \textbf{47}
(1993)~\#~5, 4474--4486.

\bibitem {MarianMarianII93}P.~Marian and T.~A.~Marian, \emph{Squeezed states
with thermal noise. II. Damping and photon counting\/}, Phys. Rev. A
\textbf{47} (1993)~\#~5, 4487--4495.

\bibitem {MisUPN13}O.~V.~Misochko, \emph{Nonclassical states of lattice
excitations: squeezed and entangled photons\/}, Phys.-Usp. \textbf{56} (2013)
\#~9, 868--882.

\bibitem {Mollow67}B.~R.~Mollow, \emph{Quantum statistics of coupled
oscillator systems\/}, Phys. Rev. \textbf{162} (1967)~\#~5, 1256--1273.

\bibitem {Mollow73}B.~R.~Mollow, \emph{Photon correlations in the parametric
frequency splitting of light\/}, Phys. Rev. A \textbf{8} (1973)~\#~5, 2684--2694.

\bibitem {Nationetal12}P.~D.~Nation, J.~R.~Johansson, M.~P.~Blencowe, and
F.~Nori, \emph{Stimulating uncertainty: Amplifying the quantum vacuum with
superconducting circuits\/}, Rev. Mod. Phys. \textbf{84} (2012),
January--March, 1--24.

\bibitem{Ni:Su:Uv} A.~F.~Nikiforov, S.~K.~Suslov, and V.~B.~Uvarov, \textsl{%
Classical Orthogonal Polynomials of a Discrete Variable\/},
Springer--Verlag, Berlin, New York, 1991.

\bibitem {Raiford70}M.~T.~Raiford, \emph{Statistical dynamics of quantum
oscillators and parametric amplification in a single mode}\textit{\/}, Phys.
Rev. A \textbf{2} (1970)~\#~4, 1541--1558.

\bibitem {Raiford74}M.~T.~Raiford, \emph{Degenerate parametric amplification
with time-dependent pump amplitude and phase}\textit{\/}, Phys. Rev. A
\textbf{9} (1974)~\#~5, 2060--2069.

\bibitem {SanSusVin}B.~Sanborn, S.~K.~Suslov, and L.~Vinet, \emph{Dynamic
invariants and Berry's phase for generalized driven harmonic oscillators\/},
Journal of Russian Laser Research \textbf{32} (2011)~\#~5, 486--494.

\bibitem {Schumaker86}B.~L.~Schumaker, \emph{Quantum mechanical pure states
with Gaussian wave functions\/}, Phys. Rep. \textbf{135} (1996)~\#~6, 318--408.

\bibitem {SchumakerCaves85}B.~L.~Schumaker and C.~M.~Caves, \emph{New
formalism for two-photon quantum optics. II. Mathematical foundation and
compact notation\/}, Phys. Rev. A \textbf{31} (1985)~\#~5, 3093--3111.

\bibitem {Scully:Zubairy97}M.~O.~Scully and M.~S.~Zubairy, \textsl{Quantum
Optics\/}, Cambridge University Press, Cambridge, 1997.

\bibitem {Stoler70}D.~Stoler, \emph{Equivalence classes of minimum uncertainty
packets}\textit{\/}, Phys. Rev. D \textbf{1} (1970)~\#~12, 3217--3219.

\bibitem {Stoler71}D.~Stoler, \emph{Equivalence classes of minimum uncertainty
packets\textit{\/. II}}\textit{\/}, Phys. Rev. D \textbf{4} (1972)~\#~6, 1925--1926.

\bibitem {Stoler74}D.~Stoler, \emph{Photon antibunching and possible ways to
observe it}\textit{\/}, Phys. Rev. Lett. \textbf{33} (1974)~\#~23, 1397--1400.

\bibitem {Suslov10}S.~K.~Suslov, \emph{Dynamical invariants for variable
quadratic Hamiltonians\/}, Physica Scripta \textbf{81} (2010)~\#5, 055006 (11~pp).

\bibitem {Suslov11}S.~K.~Suslov, \emph{On integrability of nonautonomous
nonlinear Schr\"{o}dinger equations\/}, Proc. Amer. Math. Soc. \textbf{140}
(2012)~\#~9, 3067--3082; see also arXiv:1012.3661v3 [math-ph] 16 Apr 2011.

\bibitem {VenkSatya85}M.~Venkata Satyanarayana, \emph{Generalized coherent
states and generalized squeezed coherent states}\textit{\/}, Phys. Rev. D
\textbf{32} (1985)~\#~2, 400--404.

\bibitem {VourdasWeiner87}A.~Vourdas and R.~ M.~Weiner, \emph{Photon-counting
distribution in squeezed states\/}, Phys. Rev. A \textbf{36} (1987)~\#~12, 5866--5869.

\bibitem {Walls:Milburn}D.~F.~Walls and G.~J.~Milburn, \textsl{Quantum
Optics\/}, Springer--Verlag, Berlin, Heidelberg, 2008.

\bibitem {Wilsonetal11}C.~M.~Wilson, G.~Johansson, A.~Pourkabirian, M.~Simoen,
J.~R.~Johansson, T.~Duty, F.~Nori, and P.~Delsing, \emph{Observation of the
dynamical Casimir effect in a superconducting circuit\/},~Nature \textbf{479}
(2011)~November~17, 376--379.

\bibitem {Yuen76}H.~P.~Yuen, \emph{Two-photon coherent states of the radiation
field\/}, Phys. Rev. A \textbf{13} (1976)~\#~6, 2226--2243.
\end{thebibliography}
\end{document}